\documentclass[aps,prl,reprint,superscriptaddress]{revtex4-1}
\usepackage{latexsym, amssymb, amsmath, mathrsfs, graphicx}

\begin{document}
\title{Experimental Characterization of the Stagnation Layer between Two Obliquely
Merging Supersonic Plasma Jets}

\author{E. C. Merritt}
\affiliation{Los Alamos National Laboratory, Los Alamos, NM, 87545, USA}
\affiliation{University of New Mexico, Albuquerque, NM, 87131, USA}
\author{A. L. Moser}
\affiliation{Los Alamos National Laboratory, Los Alamos, NM, 87545, USA}
\author{S. C. Hsu}
\email{Electronic address:  scotthsu@lanl.gov.}
\affiliation{Los Alamos National Laboratory, Los Alamos, NM, 87545, USA}
\author{J. Loverich}
\affiliation{Tech-X Corporation, Boulder, CO, 80303, USA}
\author{M. Gilmore}
\affiliation{University of New Mexico, Albuquerque, NM, 87131, USA}

\date{\today}

\begin{abstract}
We present spatially resolved measurements characterizing the
stagnation layer between two obliquely merging supersonic plasma
jets. Intra-jet collisionality is very high ($\lambda_{ii} \ll 1$~mm), but the inter-jet
ion--ion mean free paths are on the same order as the stagnation layer thickness (a few
cm). Fast-framing camera images show a double-peaked emission
profile transverse to the stagnation layer,
with the central emission dip consistent with a
density dip observed in the interferometer data. We demonstrate that our
observations are consistent with collisional oblique shocks.
\end{abstract}

\maketitle 

Colliding plasmas have been studied in a variety of contexts, e.g.,
counterstreaming laser-produced plasmas supporting hohlraum design for
indirect-drive inertial confinement fusion \cite{bosch,rancu,wan},
forming and studying astrophysically relevant shocks
\cite{woolsey,romagnani,kuramitsu,kugland,ross}, and for applications
such as pulsed laser deposition \cite{luna07} and laser-induced
breakdown spectroscopy \cite{sanchez-ake}. Physics issues arising in
these studies include plasma interpenetration
\cite{berger,pollaine,larroche,rambo94,rambo95,jones}, shock formation
\cite{jaffrin}, and the formation and dynamics of a stagnation layer
\cite{hough09,hough10,yeates}. In this
work, we present experimental results on two obliquely merging
supersonic plasma jets, which are in
a different and more collisional parameter regime than many of the colliding plasma examples mentioned above.
Ours and other recent jet-merging experiments
\cite{case13,messer13} were conducted to explore the feasibility of
forming imploding spherical plasma liners via an array of merging
plasma jets \cite{parks08,Awe11,davis12,cassibry12,cassibry13}, which could
have applications in forming cm-, $\mu$s-, and Mbar-scale plasmas for
fundamental high-energy-density-physics studies \cite{drake} and as a
standoff driver for magnetoinertial fusion
\cite{thio99,parks08,hsu-ieee12,santarius12,kim13}. 
Prior experiments studying
the stagnation layer between colliding laser-produced or wire-array z-pinch \cite{swadling} plasmas were on
smaller spatial scales (mm or smaller) that could not be fully
resolved by measurements. New
results in the present work are the experimental identification and characterization of a
few-cm thick stagnation layer between colliding plasmas, and the demonstration
that our observations are consistent with hydrodynamic oblique shock theory \cite{landau}.

Experiments reported here are conducted on the Plasma Liner Experiment
(PLX) \cite{hsu-pop12}, in which two supersonic argon plasma jets are
formed and launched by plasma railguns \cite{witherspoon-aps11}.
\begin{figure}[!t]
\includegraphics{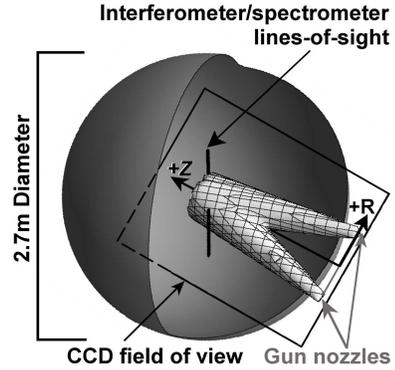}\\
\caption{Schematic of the experiment showing the spherical vacuum
  chamber, location of railgun nozzles mounted $24^\circ$ apart, two 
  merging plasma jets, $R$--$Z$ coordinates used in the
  paper, approximate interferometer/spectrometer lines-of-sight
  ($Z\approx 84$~cm), and CCD camera field-of-view.}
\label{setup}
\end{figure}
Plasma jet parameters at the exit of the railgun nozzle (peak $n_e\approx
2\times 10^{16}$~cm$^{-3}$, peak $T_e\approx 1.4$~eV, $V_{jet}\approx
30$~km/s, Mach number $M\equiv V_{jet}/C_{s,jet}\approx 14$, diameter
$= 5$~cm, and length $\approx 20$~cm) and their evolution during
subsequent jet propagation have been characterized in detail
\cite{hsu-pop12}. The jet magnetic field inside the railgun is $\sim$3~T, but the classical magnetic diffusion time is a few $\mu$s
\cite{hsu-pop12}, and thus we ignore the effects of a
magnetic field by the time of jet merging ($>20$~$\mu$s). Experimental
data are from an eight-chord laser (561~nm) interferometer
\cite{merritt-rsi12,merritt-htpd12}, a
visible-to-near-infrared 0.275~m survey spectrometer (600~lines/mm
grating and 0.45~$\mu$s gating on the 1024-pixel MCP array detector),
and an intensified charge-coupled device (CCD) visible-imaging camera (DiCam
Pro, $1280\times1024$~pixels, 12-bit dynamic range).  The
interferometer and spectrometer chords intersect the merging jets at
$Z\approx 84$~cm (Fig.~\ref{setup}).  Interferometer probe
beams (diameter $\approx 3$~mm) are each separated by 1.5~cm transverse to the merge plane. The
spectrometer has an $\approx 7$~cm diameter field-of-view in the
vicinity of the merge plane. More details about the experimental setup are given elsewhere \cite{hsu-pop12}.

\begin{figure}
\includegraphics{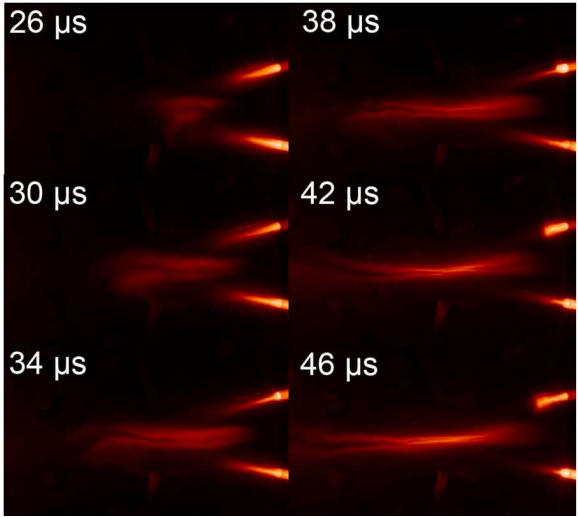}\\
\caption{False-color, slightly cropped, CCD images (log intensity, 20~ns exposure) of oblique jet merging (shots 1130, 1128, 1125, 1120, 1134, 1138). The two railgun nozzles ($\approx 46$~cm apart) are visible on the right-hand-side of each image.}
\label{pics}
\end{figure}

\begin{figure}
\includegraphics{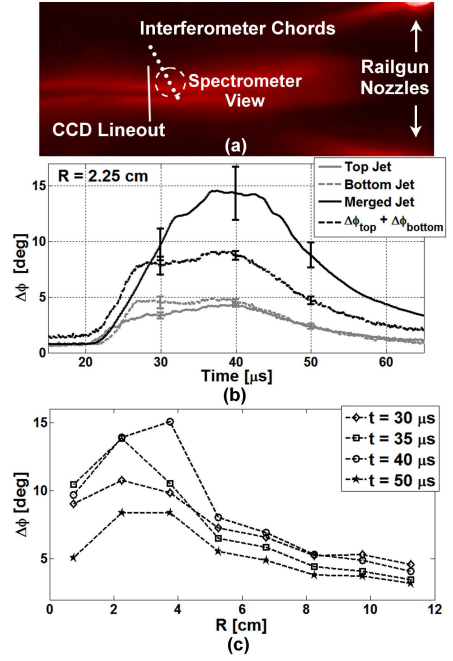}\\
\caption{(a) CCD line-out, interferometer chord, and spectrometer view positions overlayed on a CCD image ($t=38$~$\mu$s). (b)~Multi-shot-averaged interferometer phase shift vs.\ time at  $R = 2.25$~cm. (c)~Phase shift vs.\ chord position for a jet-merging case (shot 1120).}
  \label{phase}
\end{figure}

Figure~\ref{pics} shows the time evolution of oblique jet merging.
Formation of a stagnation layer between the two jets
and its double-peaked emission profile in the transverse ($R$)
direction are clearly visible. We made interferometer and spectrometer measurements
of the stagnation layer at $Z\approx 84$~cm (Fig.~\ref{phase}a) for the cases
of top-only, bottom-only, and both jets firing. Figure~\ref{phase}b
shows the interferometer phase shift $\Delta\phi$ versus time at the
$R=2.25$~cm chord position for each case. Merged-jet
measurements show that, at $R$ = 2.25 cm,
$\Delta \phi_{merge} > \Delta \phi_{top} + \Delta\phi_{bottom}$ (Fig.~\ref{phase}b), implying that simple jet
interpenetration cannot account for the observed $\Delta\phi$ of the
merged-jet stagnation layer (more quantitative analysis given later).
However, at large $R$ (e.g., 6.75 cm), $\Delta \phi_{merge}
\approx \Delta \phi_{top} + \Delta \phi_{bottom}$ (not shown), consistent with
jet interpenetration. Figure~\ref{phase}c shows
$\Delta\phi$ vs.\ $R$ at four times for a merged jet.
The $\Delta\phi$ dip at $R=0.75$~cm and peak at $R=2.25$--4~cm are
well-aligned with the emission dip and peak (Fig.~\ref{phase}a), respectively.

The interferometer $\Delta\phi$ measurements are used to estimate the
ion plus neutral density $n_{tot}$. Our interferometer is
sensitive to bound and free electrons in the plasma
\cite{merritt-htpd12}, so $\Delta \phi$ contributions from all species
and ionization states must be considered. The $\Delta \phi$ satisfies
$\int
n_{tot} dl = \frac{\Delta \phi}{C_e [Z_{\rm eff} - Err]}$, where the integral is over
the chord length, $C_e =
\lambda e^2/ 4 \pi \epsilon_0 m_e c^2$ is the phase sensitivity to
electrons, $Z_{\rm eff} = n_e/ n_{tot}$ is the mean charge,
$Err = \sum_{j,k} (2 \pi/\lambda C_e n_{tot}) K_{j,k} m_k n_{j,k}$ is
the error in the phase shift due to all ionization states $j$ of all
ion species $k$, $m_k$ is the atomic mass of ion species $k$, and $K_{j,k}$ is the specific
refractivity. Uncertainty in plasma jet composition (due to impurities)
can be accounted for by bounding $n_{tot}$ at the extremes  $Err = 0$
and $Err = Err_{max} = (2 \pi m/\lambda C_e ) K_{max}$, where
$K_{max}$ is the largest specific refractivity of all the species
present. Since we do not know precisely the impurity fraction or mixture
ratios of impurities, we perform our data analysis by
considering the two extreme
cases of (i)~100\% argon and (ii)~30\% argon with 70\% impurities. The latter is chosen based on
the difference in measured chamber pressure rise for gas-injection-only versus full plasma discharges. Identification of bright Al and O spectral lines in our data suggest that impurities are from the alumina-based railgun insulators. Thus, for case (ii), we approximate the plasma jet to be 43\% O and 24\% Al and assume that $Err_{max} = Err_{\rm Al \textsc{i}}$. Top-jet-only experiments provide single-jet $\Delta \phi$ vs.\ time at $Z \approx 84$~cm and $R = 2.25$~cm (Fig.~\ref{phase}b). The average single-jet peak phase shift is $\Delta \phi =4.3
\pm 0.3^\circ$ for the data considered (shots 1265--1267). All chord positions $R = 0.75$--11.25~cm (Fig.~\ref{phase}a) have similar $\Delta\phi \approx 4^\circ$. Using $\Delta \phi = 4^\circ$, $Z_{\rm eff}$ = 0.94 (inferred from spectroscopy analysis
\cite{hsu-pop12} assuming 100\% argon), $Err_{max} = 0.082$, and a jet diameter of 22 cm (from
CCD images \cite{hsu-pop12}), gives a single-jet density range of
$n_{tot} = n_{single}=2.1$--$2.3 \times 10^{14}$~cm$^{-3}$.  The latter result changes
by only a few percent for the 30\%/70\% mixture, which
changes the inferred $Z_{\rm eff}$ to 0.92.

\begin{figure}
\includegraphics{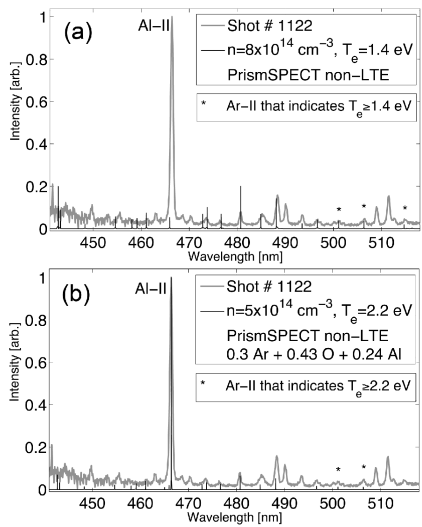}\\
\caption{Spectral data (gray) and non-LTE PrismSPECT calculations
  (black) for the merged-jet stagnation layer ($Z \approx 84$~cm, $t =
  36$~$\mu$s).  The PrismSPECT calculations are for (a)~100\% argon
  and (b)~30\%/70\% mixture.  Lower bounds on peak $T_e$ are
  inferred based on the presence of the Ar~\textsc{ii} lines indicated
  by asterisks.}
\label{spectra}
\end{figure}

By using the interferometry estimates for merged-jet density (shown later) and comparing our spectroscopy data with non-LTE spectral calculations using PrismSPECT \cite{macfarlane03}, we
infer $Z_{\rm eff}$ and $T_e$ of the stagnation layer at
$Z\approx 84$~cm. PrismSPECT results for $Z_{\rm eff}$ and $T_e$ are
sensitive to the specific plasma mixture used.
Based on the presence of certain Ar~\textsc{ii} lines in the data and by comparing to
PrismSPECT results, we bound estimates of $Z_{\rm eff}$ and $T_e$ using the 100\% argon and 30\%/70\%
mixture cases. For the former (Fig.~\ref{spectra}a), we infer that
peak $T_e \ge 1.4$~eV and $Z_{\rm eff} = 0.94$.
For the latter (Fig.~\ref{spectra}b), we infer that
$2.2$~eV$\le$ peak $T_e < 2.3$~eV and $Z_{\rm eff}=1.3$--1.4, with the
upper bounds determined by the absence of an Al~\textsc{iii} line in
the data. For the mixture, $M\approx 9$ compared to $M\approx
14$ for pure argon.

With estimates of the stagnation layer $Z_{\rm eff}$ in hand, we estimate
the stagnation layer density and compare it with the single-jet
(un-shocked) density. At $R$ = 2.25 cm, the average peak $\Delta \phi =
14.3 \pm 2.4^\circ$ (Fig.~\ref{phase}b) (shots 1117--1196). Using $\Delta \phi = 14^\circ$, chord path length of
22~cm, and $Z_{\rm eff}$ = 0.94 (100\% argon case), $n_{tot} = n_{merged}= 7.5$--$8.2
\times 10^{14}$~cm$^{-3}$. In this case the density increase
$n_{merged}/n_{single} = 3.2$--3.8. For $Z_{\rm eff} = 1.4$ (30\%/70\% mixture case), the stagnation layer density is $n_{tot} = n_{merged}=
5.0$--$5.3 \times 10^{14}$~cm$^{-3}$, and the density increase is
$n_{merged}/n_{single} = 2.1$--2.5.  Thus, the observed range of $n_{merged}/n_{single}=2.1$--3.8,
exceeding the factor of two expected for jet interpenetration.

We compare the experimentally inferred
density jumps with oblique shock theory \cite{landau}. At $Z \approx
84$~cm and $M = 9$--14, the theory predicts a shock angle $\beta
\approx 19^\circ$--$20^\circ$, as discussed later. For $\gamma = 1.4$, the
predicted density jump across an oblique shock \cite{drake} is $n_{shock}/n_{unshocked} = (M \sin \beta)^2 (\gamma + 1)/[(M \sin \beta)^2 (\gamma - 1) -2] \approx 4.0$--4.9.
Difference between the measured and
predicted density jumps could be due to 3D (e.g.,
pressure-relief in the out-of-page dimension) and/or equation-of-state (e.g., ionization \cite{messer13})
effects not modeled by the theory.

\begin{figure}
  \includegraphics{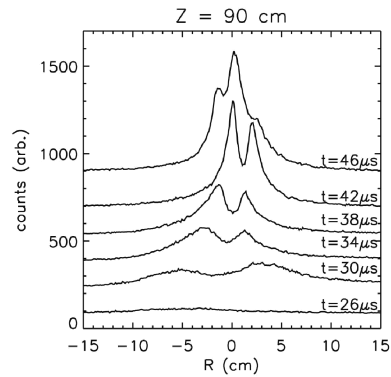}\\
  \caption{CCD image line-outs versus $R$ (transverse to stagnation layer) at
$Z=90$~cm (horizontal pixel \# 654 out of 1024), corresponding to the images of
Fig.~\ref{pics}. Progressive times are shown with increasing count offsets to avoid
trace overlap.}
\label{lineout}
\end{figure}

The stagnation layer thicknesses as observed in the merged-jet emission
(Fig.~\ref{lineout}) and $\Delta \phi$ vs.\ $R$ profiles (Fig.~\ref{phase}c) are similar in scale (few cm).
In a collisional plasma, the layer thickness
is expected \cite{rambo94} to be of order the counter-streaming ion--ion mean
free path (mfp) $\lambda_{ii^\prime}\sim v_{rel}/4\nu_{ii^\prime}$ \cite{messer13}, where
$v_{rel}$ is the relative transverse velocity between obliquely merging jets and
the slowing-down rate in the fast approximation ($v_{rel}\gg v_{ti}$) is
$\nu_{ii^\prime}\approx 9.0\times 10^{-8} (1/\mu+1/\mu^\prime)(\mu^{1/2}/\epsilon^{3/2})
n_{i^\prime}(Z_{\rm eff}Z^\prime_{\rm eff})^2 \ln \Lambda$
\cite{nrl-formulary}.  Note that in our parameter regime, the inter-jet $\lambda_{ie} \gtrsim \lambda_{ii}$.
We estimate
$\lambda_{ii^\prime}$ by considering jets of 100\% argon
and the 30\%/70\% mixture previously discussed, in all cases
using $v_{rel}=20$~km/s.
For Ar--Ar stopping, $\lambda_{ii}=3.47$~cm (for
$n_i = 8 \times 10^{14}$~cm$^{-3}$,
$T_e = 1.4$~eV, $Z_{\rm eff} = 0.94$). Pure Al--Al and O--O stopping yield
$\lambda_{ii}=0.16$ and 0.62~cm, respectively (for $n_i=5\times10^{14}$~cm$^{-3}$,
$T_e = 2.2$~eV, $Z_{\rm eff,Al} =
2.0$, $Z_{\rm eff,O} = 1.0$).  For inter-species collisions in a
mixed-species jet, using the mixture given in Fig.~\ref{spectra}b,
$\lambda_{ii^\prime} \approx 0.57$--6.18~cm.  We have also
estimated the inter-jet mfp due to Ar$^+$--Ar charge-exchange and momentum
transfer \cite{phelps90} to be $\approx 3$~cm.  These estimates
imply that our inter-jet merging is collisional, which is consistent with
a more detailed treatment of inter-jet ion--ion
stopping including jet profile effects \cite{messer13}.

\begin{figure}[!t]
  \includegraphics{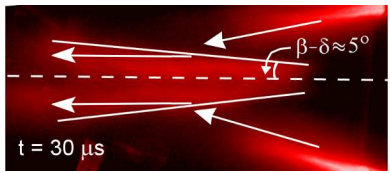}\\
  \caption{CCD image (shot 1089, $t = 30$~$\mu$s) with postulated shock boundaries
(solid white lines) and observed shock angle $\beta - \delta \approx 5^\circ$
relative to the midplane.}
  \label{boundary}
\end{figure}

Assuming that the emission layers are post-shocked plasma, we postulate that their edges (at larger $|R|$) are the shock boundaries (Fig.~\ref{boundary}). Qualitatively, the merging geometry for an individual jet resembles that of a supersonic flow past a wedge \cite{landau} and should result in the formation of a linear attached shock at an angle $\beta$ with respect to the original flow direction. The angle $\beta$ is a function of $M$ and the wedge angle $\delta$ between the jet flow and the midplane, and satisfies $\tan \delta = 2 \cot \beta [\frac{M_1^2 \sin^2 \beta - 1}{M_1^2( \gamma + \cos 2\beta) + 2}]$ \cite{landau}.  In our case, $\tan \delta = (23$~cm)$/Z_i$, where $Z_i$ is the point at which jets first interact. We estimated $Z_i$ from CCD images as the minimum $Z$ for which merged-jet emission is observed, with $Z_i \approx 45$~cm at $t = 26$~$\mu$s evolving to $Z_i \approx 25$~cm at $t = 46$~$\mu$s. For this range of $Z_i$, we predict shock angles (relative to the midplane) $\beta - \delta \approx 8^\circ$--$17^\circ$ ($M = 9$) and $7^\circ$--$15^\circ$ ($M = 14$). For shot 1089 with $Z_i \approx 30$~cm, $\beta - \delta \approx 5^\circ$ (Fig.~\ref{boundary}), which is within a factor of two of the predicted $\beta - \delta \approx 9^\circ$--$10^\circ$. This is reasonable given that the prediction does not include 3D nor equation-of-state \cite{messer13} effects. Oblique shock theory also predicts the formation of a detached shock \cite{landau}, which would have a curved shock boundary for $\delta > \delta_{max} \approx 45^\circ$ (for $M=9$--14 and $Z_i\approx 25$~cm at late times). CCD images at $t \ge 42$~$\mu$s (Fig.~\ref{pics}) show curvature of the emission layers away from the midplane, suggestive of detached shocks.

\begin{figure}[!b]
  \includegraphics{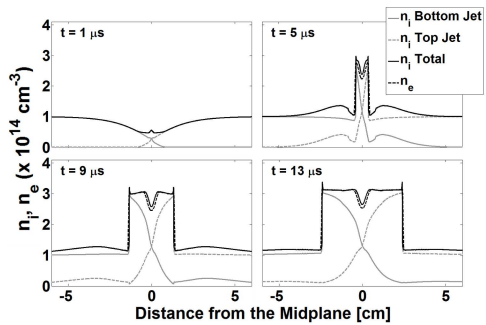}\\
  \caption{Density profiles from a 1D multi-fluid
    collisional plasma
    simulation that models the transverse ($R$) dynamics of our oblique jet-merging experiments.}
  \label{sim}
\end{figure}

To further evaluate the 
consistency of the observed stagnation layer structure with hydrodynamic
oblique shock theory \cite{landau}, and to help interpret the emission profile dips seen in Figs.~\ref{pics} and \ref{lineout}, we model the transverse ($R$) dynamics of the oblique merging
using 1D multi-fluid collisional plasma simulations of merging jets.  We treat
the electrons as one fluid and the ions of each jet as a second and third fluid, respectively, thus allowing for interpenetration between the two jets.  Simulations were performed using the 
USim code (formerly called Nautilus) \cite{loverich,loverich13}.  Algorithms on which USim is based have been verified against shock-relevant problems \cite{loverich06,loverich11}. In the simulations, the jets are assumed to be 100\% Ar~\textsc{ii} with initial $n_e = n_i = 10^{14}$~cm$^{-3}$, $T_e = T_i = 1.4$~eV, and velocities of $\pm 6.2$~km/s (i.e., transverse component of $V_{jet}\approx 30$~km/s).  We used a density profile in the
leading edge of the jet as shown in the upper-left panel of
Fig.~\ref{sim}.  The simulation used a cell size of $\sim 100$~$\mu$m. 
Details of the semi-implicit numerical algorithm are described in \cite{kumar2012}. 
As shown in Fig.~\ref{sim}, at 1~$\mu$s after merging begins,
there is a small initial
density buildup at the midplane; the electrons are very highly collisional and thus the incoming electron fluid of each jet must pile up at the midplane.  At 5 $\mu$s, outward-propagating, sharp density jumps have formed, and a density dip appears at the midplane.  The density jumps are consistent with reflected shocks in that the density jump ($\approx  3\times$) and its propagation speed ($\approx 2.4$~km/s), as observed in the simulation, agree very well (within 5\%) with Rankine--Hugoniot jump
condition predictions using the upstream and downstream densities and pressures.  The midplane density dip (consistent
with the emission profile dip) 
arises to maintain pressure balance in the presence of midplane shock heating.
The relatively slow reflected shock speed is consistent with the experimental observation in that the emission peaks do not move very far over $>10$~$\mu$s. Finally, interpenetration between the two jets reaches $\sim 1$~cm (lower-right panel of Fig.~\ref{sim}), consistent with earlier estimates of inter-jet ion--ion collisional mfp $\sim 1$~cm.  These comparisons support the interpretation that our experimental
observations are consistent with collisional oblique shocks.

In summary, we have experimentally characterized the stagnation layer
between two obliquely merging supersonic plasma jets. The jets are
individually very highly collisional, but the inter-jet ion--ion collisional mfp is of order the
stagnation layer thickness of a few centimeters. CCD images show the
formation of a stagnation layer with a double-peaked emission
profile transverse to the layer, with the central emission dip
consistent with a density dip observed in the interferometer data. The
geometry of the observed stagnation layer structure is
consistent with hydrodynamic oblique shock theory. Furthermore,
collisional 1D multi-fluid plasma simulations that model the
transverse dynamics of the oblique merging show the formation and
evolution of reflected shocks with a central density dip consistent
with the observed stagnation layer emission profile dip.  Ongoing experiments are now
employing lower jet density and higher jet velocity to study head-on
jet merging with very low inter-jet ion--ion collisionality.

\begin{acknowledgments}
We acknowledge J. Dunn for experimental support, S. Brockington and
F. D. Witherspoon for advice on railgun operation, and C. Thoma and
J. Cassibry for discussions regarding the simulations. This work was
supported by the U.S. Dept.\ of Energy.
\end{acknowledgments}

%


\end{document}